\documentclass[aps,pra,superscriptaddress,twocolumn,nofootinbib]{revtex4-1}

\usepackage{graphicx}   
\usepackage{amsmath,amsfonts,amssymb,color,bbm, braket}
\usepackage{enumerate, relsize}

\usepackage[ocgcolorlinks,colorlinks=true,linkcolor=blue,citecolor=red]{hyperref}
\usepackage{breakurl}

\newcommand{\Mod}[1]{\left|#1\right|}

\newcommand{\tr}{\mathrm{tr}}

\newcommand{\ketbra}[2]{|#1\rangle\langle#2|}

\newcommand{\ew}{{\mathlarger{\eta}_{\mathsmaller{W}}}}
\newcommand{\et}[4]{\mathlarger{\eta}^{\mathsmaller{#1\,#2}}_{\mathsmaller{#3\,#4}}}
\newcommand{\ett}[5]{\mathlarger{#5}^{\mathsmaller{#1\,#2}}_{\mathsmaller{#3\,#4}}}
\newcommand{\chan}[3]{\mathlarger{#1}^{\mathsmaller{#3}}_{\mathsmaller{#2}}}

\newcommand{\vA}{\mathcal{A}}
\newcommand{\vB}{\mathcal{B}}
\newcommand{\vC}{\mathcal{C}}
\newcommand{\oA}{A}
\newcommand{\oB}{B}

\begin{document}

\author{Ralph Silva}\affiliation{Group of Applied Physics, University of Geneva, CH-1211 Geneva 4, Switzerland}
\author{Yelena Guryanova}\affiliation{Institute for Quantum Optics and Quantum Information (IQOQI), Boltzmanngasse 3 1090, Vienna, Austria}
\author{Anthony J. Short}\affiliation{H. H. Wills Physics Laboratory, University of Bristol, Tyndall Avenue, Bristol, BS8 1TL, United Kingdom}
\author{Paul Skrzypczyk}\affiliation{H. H. Wills Physics Laboratory, University of Bristol, Tyndall Avenue, Bristol, BS8 1TL, United Kingdom}
\author{Nicolas Brunner}\affiliation{Group of Applied Physics, University of Geneva, CH-1211 Geneva 4, Switzerland}
\author{Sandu Popescu}\affiliation{H. H. Wills Physics Laboratory, University of Bristol, Tyndall Avenue, Bristol, BS8 1TL, United Kingdom}

\title{Connecting processes with indefinite causal order and multi-time quantum states}

\begin{abstract}
Recently, the possible existence of quantum processes with indefinite causal order has been extensively discussed, in particular using the formalism of process matrices. Here we give a new perspective on this question, by establishing a direct connection to the theory of multi-time quantum states. Specifically, we show that process matrices are equivalent to a particular class of pre- and post- selected quantum states. This offers a new conceptual point of view to the nature of process matrices. Our results also provide an explicit recipe to experimentally implement any process matrix in a probabilistic way, and allow us to generalize some of the previously known properties of process matrices. Furthermore we raise the issue of the difference between the notions of indefinite temporal order and indefinite causal order, and show that one can have indefinite causal order even with definite temporal order. 
\end{abstract}
\maketitle

\section{Introduction}

When describing physical phenomena it is commonly assumed that there exists an underlying causal order. Loosely speaking, later events can be influenced by previous ones, but not the other way around. Recently, however, the idea that physical theories necessarily require a causal order has been challenged, in particular in the context of quantum theory. Indeed, one may imagine that the notion of causal order can be subject to fundamental quantum principles, such as the superposition or uncertainty principle, resulting in indefinite causal structures. Ref. \cite{Aharonov90} introduced the notions of a ``superposition of quantum evolutions'' and of what is now known as a ``quantum switch'' \cite{giulioIndefinite}. The question of the existence of quantum correlations with indefinite causal order has recently triggered an intense research effort and subsequently several frameworks for characterizing quantum processes with undefined or dynamical causal structures have been developed, see e.g. \cite{giulioIndefinite,Hardy07,Leifer13,Baumeler14,Dominic16}. 

In \cite{Oreshkov2012}, Oreshkov, Costa, and Brukner introduced the  framework of `process matrices' where causal structure can be partially relaxed. Here, operations in local laboratories are described by quantum theory (making use of a local causal structure), however, no global causal structure is assumed. Crucially, this framework captures situations that cannot be explained by any classical causal structure, as witnessed by the violation of so called ``causal inequalities'' (analogous to Bell inequalities) \cite{Oreshkov2012,Branciard16a,Araujo1,Oreshkov16,Abbott16,Baumeler16}. Possible implications for quantum information processing have also been discussed \cite{Chiribella12,Araujo14}.

Although the process matrix formalism contains features reminiscent of quantum mechanics, it is not derived from quantum theory. The crucial question of its relation to quantum theory was left open in the original work. It could have been the case that process matrices simply represented a reformulation of some specific situations in quantum mechanics, or that they contained new physics outside of quantum theory. 

Here we offer an answer to this question. We show that process matrices correspond to  a subclass of ``two-time states", which were introduced by Aharonov and his collaborators \cite{ABL, Aharonov1991, Ralph2014}. This means that the world described by process matrices is equivalent to a particular case of a quantum world with {\it fundamental} post-selection, i.e. a quantum world with independent initial and final boundary conditions, which are both guaranteed to occur (as opposed to ordinary post-selection in which the final state occurs only probabilistically). 

In \cite{Oreshkov16} an alternative connection between quantum mechanics and process matrices was found, via a powerful generalisation of the process matrix formalism.  In particular, as a corollary, it was shown that any process matrix can be simulated in quantum mechanics using post-selection. As any two-time state can be probabilistically implemented experimentally in standard quantum theory via post-selection, we also recover this result in our formalism.

A number of further insights also follow from our results. First, and most importantly, they show that there is a distinction between indefinite temporal order and indefinite causal order, and that one can have indefinite causal order with definite temporal order. Furthermore, they allow us to generalise some of the previously known properties of process matrices, which might be of interest in future work.

Finally, we show that the subset of pre- and post-selected states corresponding to process matrices has very special properties, leading to probabilities that are linear functions of the states and of the measurements. This raises a new and important question: Why only this subset? What would be problematic if the probabilities were non-linear functions, as is the case for most pre- and post-selected states?

\section{Pre- and post-selection: experimental versus fundamental}
The standard procedure for collecting statistics in quantum mechanics is to fix an initial state for a system and accumulate experimental results based on this chosen state, or perhaps on an evolution of it later in time. 
As pointed out by Aharonov et al. in \cite{ABL}, in addition to fixing an initial state for a system, one may also specify an independent final state. One way to realise such a specification is with post-selection, which can be understood by considering the following scenario.

At the initial time $t_1$, Alice prepares a quantum system in the state $\ket{\psi}$. In the time interval between the initial and a final time $t_2$ she performs some arbitrary experiments and records their results. At the final time $t_2$, she measures an observable $O$, one of whose non-degenerate eigenstates is $\ket{\phi}$, the desired final state (which is arbitrary and independent of $\ket{\psi}$). Alice considers her experiment to be successful if the measurement of $O$ yields the eigenvalue corresponding to the eigenstate $\ket{\phi}$; otherwise she discards the experiment.

This way, if Alice repeats her experiment on an ensemble of particles, all prepared in the initial state $\ket{\psi}$, she ends up with a sub-ensemble, which we call a pre- and post-selected ensemble, characterised by the initial state $\ket{\psi}$ and the final state $\ket{\phi}$. In this sub-ensemble the statistics of the results of the intermediate measurements is, in general, different from the statistics over the entire ensemble. 

Importantly, the procedure described above is purely quantum mechanical, albeit not the one that is usually considered (the standard paradigm considers only pre-selected ensembles).
In this procedure, one cannot guarantee a priori that the final measurement will yield the eigenvalue corresponding to $\ket{\phi}$. Thus, at no intermediate point can one know if the post-selection will be successful or not (Alice may even decide not to measure $O$). It follows that there is also no intermediate time at which Alice can know that she is in the desired pre- and post-selected sub-ensemble. Only after the final time, $t_2$, will she know which events are successful and must be kept, and which have to be discarded. It is only then, looking back at the records of her intermediate measurement results, that she is be able to find out the statistics corresponding to a specific pre- and post-selected ensemble. 

Contrary to this experimental realisation of a pre- and post-selected ensemble, the quantum mechanical formalism also allows one to fix a guaranteed final state. { Unlike the case of experimental post-selection, here} Alice would already see the statistics corresponding to the pre- and post-selected ensemble at the intermediate times $t_1 < t< t_2$. We call this a situation with `fundamental' post-selection, as opposed to the ordinary, measurement based, probabilistic post-selection. 

There is no evidence that fundamental post-selection exists in nature, however, some authors have suggested that it could be present in some exotic situations, for example having a final state of the universe \cite{AhaPeronal,*AhaRoh05,*Aharonov,Hartle,Gell-Mann}, or a final state at the singularity of a black-hole \cite{maldacena}.

\section{The Two-time state formalism}
Regardless of whether a two-time state arises from experimental or fundamental post-selection, the statistics generated by a given pre- and post-selected ensemble are the same. First introduced for pure states \cite{ABL,Aharonov1991}, the formalism was recently extended to general mixtures of two-time states \cite{Ralph2014}. In this section we will review this formalism following the convenient notations of Ref. \cite{Ralph2014}.\\
\indent
Suppose, once more, that Alice starts at $t_1 $with her system in the state $\ket{\psi}$, measures the operator $O$ and selects the cases in which this final measurement yields the eigenvalue corresponding to $\ket{\phi}$ at time $t_2$. For simplicity we assume that the Hamiltonian between $t_1$ and $t_2$ is zero and that in the intermediate time Alice performs a detailed measurement, described by the set of Kraus operators $\{\hat{E}_a = \sum_{k,l} \beta_{a,kl} \ket{k}\bra{l} \}$. By ``detailed measurement'' we mean a measurement where each outcome $a$ corresponds to a single Kraus operator, such that the normalisation condition that the operators obey is $ \sum_a \hat{E}^\dagger_a \hat{E}_a = \mathbb{I}$. The probability to obtain the outcome $a$ given the pre- and post-selection is then given by  

\begin{equation}\label{eq:pureProb}
P(a) = \frac{
	\Mod{\braket{\phi|\hat{E}_a|\psi}}^2}
		{\sum_{a^\prime} \Mod{
			\braket{\phi|\hat{E}_{a^\prime}|\psi}
				}^2
			}.
\end{equation}
In order to use the two-time formalism to its full advantage, we transfer to the two-time language. Formally,  we take the state space of Alice to be the tensor Hilbert space $\mathcal{H}_{\vA_2} \otimes \mathcal{H}^{\vA_1}$, where $\mathcal{H}^{\vA_1}$ is the Hilbert space of the pre-selected  states, which are denoted by ket vectors (with raised labels) and that evolve forward in time, and $\mathcal{H}_{\vA_2}$ is the Hilbert space of the post-selected states, which are denoted by bra vectors (with lowered labels) and that evolve backward in time. The structure is made explicit by defining a {two-time} state and a two-time version of the Kraus operator in the following way
\begin{align}\label{eq:exampleone}
\Psi &= {}_{\vA_2}\!\bra{\phi} 
	\otimes 
		\ket{\psi}^{\vA_1}\;\;& 
			\in \mathcal{H}_{\vA_2} 
				\otimes \mathcal{H}^{\vA_1}, \\
E_a &= 
	\sum_{kl} 
		\beta_{a,kl}
			\ket{k}^{\vA_2} \otimes
				 {}_{\vA_1}\!\bra{l} \;\; 
				 	&\in \mathcal{H}^{\vA_2}
				 		 \otimes 
				 		 	\mathcal{H}_{\vA_1}.
\end{align}
where we differentiate between the usual Kraus operator and its two-time version by the presence or absence of a hat.

This notation places states and measurement operators on an equal footing such that they are dual to one another, which allows us to re-write the amplitude $\braket{\phi|\hat{E}_a|\psi}$ appearing in \eqref{eq:pureProb} as  $\Psi \bullet E_a$. The  operation $(\bullet)$ applies to vectors belonging to different Hilbert spaces and is a combined composition/contraction: it contracts any dual vector pairs (i.e. bra and ket pairs) in Hilbert spaces with the same labels to generate a scalar\footnote{Note that we follow the convention from general relativity -- a contraction can only occur between a raised and a lowered label.}, e.g. ${}_{\vA_2}\bra{\phi}\bullet\ket{i}^{\vA_2} = \braket{\phi|i}$; and it performs the tensor product between any unpaired vectors. In the present example all the vectors are paired, so $\bullet$ effectively results in a scalar product; the tensoring will be useful in later examples.\footnote{Since in this framework every Hilbert space (bra or ket) has a distinct label, the order in which the Hilbert spaces are written is arbitrary.}

The probability to obtain the outcome $a$, as seen in \eqref{eq:pureProb}, now takes on the alternative form
\begin{equation}\label{purestateprob}
P(a) = \frac{\Mod{\Psi\bullet E_a}^2}{\sum_{a^\prime} \Mod{\Psi \bullet E_{a^\prime}}^2}.
\end{equation}

\begin{figure}[t!]
	\includegraphics[width=0.9\linewidth]{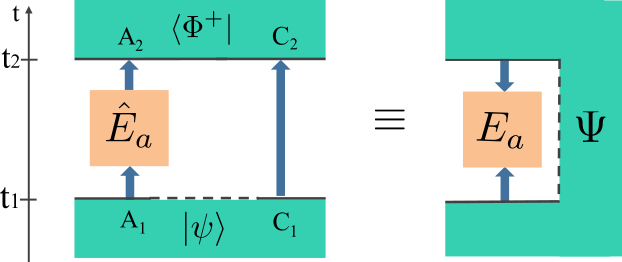}
	\caption{ 
Representation of how to experimentally prepare an entangled two-time state. An experimenter prepares the state $\sum_{ij} \alpha_{ij} \ket{j}_{A_1} \otimes \ket{i}_{C_1}$ of Alice's system and an ancilla (denoted $C$). After Alice has performed her operation, the experimenter then post-selects the final state of Alice and the ancilla to be the maximally entangled state $(\sqrt{d})^{-1} \sum_k \ket{k}_{A_2} \otimes \ket{k}_{C_2}$, where $d$ indicates the dimension of the space on $A_2$ and also that of the ancilla. Note that this procedure can be intuitively understood as entanglement swapping: the entanglement between the system and ancilla is effectively swapped to the two-time state of the system. One can verify that the statistics of the outcomes of Alice's operation in this case obey \eqref{purestateprob}.}
\label{fig:prepareState}
\end{figure}

The advantage of viewing pre- and post-selections as two-time states in the tensor product Hilbert space $ \mathcal{H}_{\vA_2} \otimes \mathcal{H}^{\vA_1}$ is that one can take superpositions of simple ``direct product" two-time states and obtain general pure two-time states in which the pre-selection is entangled with the post-selection. An arbitrary pure two-time state has the form
\begin{equation}\label{examplepurestate}
\Psi = \sum_{ij} \alpha_{ij} \; {}_{\vA_2}\!\bra{i} \otimes \ket{j}^{\vA_1} \;\;\;\;\;\in \mathcal{H}_{\vA_2} \otimes \mathcal{H}^{\vA_1}. 
\end{equation} 
This general two-time state can be understood as a fundamentally post-selected state, or one that can be implemented via experimental post-selection using entangled ancillas \cite{multi}, the procedure for which is detailed in Fig.~\ref{fig:prepareState}.

Generalizing further, one may also consider mixtures of pure two-time states \cite{Ralph2014}. Consider that the experimenter follows a similar procedure as above (see Fig.~\ref{fig:prepareState}). However, instead of preparing a pure state for $A_1$ and the ancilla $C$, the experimenter prepares any possible mixed state, i.e. the ensemble $\{\sum_{ij} \alpha_{r,ij} \ket{j}_{A_1} \otimes \ket{i}_{C}\}_r$ with associated probabilities $p_r$.

In this case, the statistics of Alice's operation obey
\begin{equation}\label{e:probs 1}
P(a) = \frac{\sum_r p_r \Mod{\Psi_r \bullet E_a}^2}{\sum_{a^\prime} \sum_{r^\prime} p_{r^\prime} \Mod{\Psi_{r^\prime} \bullet E_{a^\prime}}^2}.
\end{equation}
where $\Psi_r = \sum_{ij} \alpha_{r,ij} \; {}_{\vA_2}\!\bra{i} \otimes \ket{j}^{\vA_1}$

Analogous to the case of ordinary quantum mechanics where the density operator captures the information of a mixed state at a single time, one can also construct a similar object for multi-time states. The \textit{density vector} of a pure multi-time state $\Psi$ is given by $\Psi \otimes \Psi^\dagger$, where the Hilbert spaces pertaining to $\Psi^\dagger$ are differentiated by dagger labels, such that every ket (bra) in $\Psi$ is transformed into a daggered bra (ket) in $\Psi^\dagger$.\footnote{Note that, given our convention that ket vectors have raised labels and bra vectors lowered labels, the dagger operation also swaps the position of the label.}

For the pure state in \eqref{examplepurestate},
\begin{equation}
	\begin{split}
	\Psi \otimes \Psi^\dagger &\in \mathcal{H}_{\vA_2} \otimes \mathcal{H}^{\vA_1} \otimes \mathcal{H}_{\vA_1^\dagger} \otimes \mathcal{H}^{\vA_2^\dagger}, \\
	\Psi \otimes \Psi^\dagger &= \sum_{ijmn} \alpha_{ij} \alpha_{mn}^* \;
	{}_{\vA_2}\!\bra{i} \otimes \ket{j}^{\vA_1} \otimes {}_{\vA_1^\dagger}\!\bra{n} \otimes \ket{m}^{\vA_2^\dagger}
	\end{split}
\end{equation}
We will use the shorthand $\mathcal{H}^{\oA} := \mathcal{H}^{\vA}\otimes \mathcal{H}_{\vA^\dagger}$ and $\mathcal{H}_{\oA} := \mathcal{H}_{\vA}\otimes \mathcal{H}^{\vA^\dagger}$ from here on, such that $\Psi \otimes \Psi^\dagger \in \mathcal{H}^{\oA_1}\otimes \mathcal{H}_{\oA_2}$.

The density vector of the ensemble is the convex combination of the pure density vectors,
\begin{equation}\label{e:eta ensemble}
\eta = \sum_r p_r \Psi_r \otimes \Psi_r^\dagger \quad \in \mathcal{H}^{\oA_1}\otimes \mathcal{H}_{\oA_2}.
\end{equation}

Similar to the construction of a density vector for states, one can also construct a `Kraus density vector' $J_a = E_a \otimes E_a^\dagger$ for any Kraus operator $\hat{E}_a$. If the measurement is not detailed with respect to the outcome $a$ (in other words multiple Kraus operators correspond to a single outcome, a so-called  `coarse-grained' measurement), then the density vector corresponding to the outcome is the sum of all the Kraus density vectors corresponding to that outcome. If $\hat{E}_{a}^\mu$ denotes the Kraus operator, where $a$ is the outcome and $\mu$ is an index running over all the operators corresponding to that outcome,

\begin{equation}\label{Krausdensityvector}
J_a = \sum_\mu E_a^\mu \otimes {E_{a}^\mu}^\dagger \quad \in \mathcal{H}_{\oA_1} \otimes \mathcal{H}^{\oA_2}.
\end{equation}
It is important to note that $J_a$ contains all the information about the dynamics induced by the measurement, and not only information about the outcome probabilities. That is, it is not equivalent to the POVM element $\sum_\mu \hat{E}_a^{\mu\dagger}\hat{E}_a^\mu$, which contains no information about the post-measurement state.

The normalisation condition in the standard formalism, $\sum_{a,\mu} \hat{E}_a^{\mu\dagger} \hat{E}_a^\mu = \mathbb{I}$, associated to the fact that the Kraus operators $\{\hat{E}_a^\mu\}_{a,\mu}$ form a completely-positive trace-preserving channel, expressed in our notation is\footnote{This representation is the analogue of the condition that if $\Lambda(\cdot) = \sum_{a,\mu}\hat{E}_a^\mu (\cdot)\hat{E}_a^{\mu\dagger}$ is a trace-preserving quantum channel, then the conjugate channel $\Lambda^\dagger(\cdot) = \sum_{a,\mu}\hat{E}_a^{\mu\dagger} (\cdot){\hat{E}_a^\mu}$ is \emph{unital}, i.e.~ satisifes $\Lambda^\dagger(\mathbb{I}) = \mathbb{I}$.}
\begin{equation}\label{e:future id pres}
\chan{\mathbb{I}}{\oA_2}{}\bullet \chan{J}{\oA_1}{\oA_2} = \chan{\mathbb{I}}{\oA_1}{},
\end{equation} 
where $J = \sum_a J_a$ and we have introduced the `identity' vector $\chan{\mathbb{I}}{\oA}{} := \chan{\mathbb{I}}{\vA}{\vA^\dagger} \in \mathcal{H}_\oA$ given by
\begin{equation}\label{e:identity op}
\chan{\mathbb{I}}{\vA}{\vA'} = \sum_i \ket{i}^{\vA'}\otimes{}_{\vA}\bra{i} \quad \in \mathcal{H}^{\vA'} \otimes \mathcal{H}_{\vA}.
\end{equation}
Considering the form of \eqref{e:future id pres}, we can think of $J$ that satisfy this condition as \emph{future identity preserving}.

The generalization to mixed states and non-detailed measurements preserves the duality between states and measurements, that is now between the density vectors corresponding to two-time states and the density vectors corresponding to a measurement outcome. The statistics of Alice's operation given by \eqref{e:probs 1} can now be concisely written as
\begin{equation}\label{Born2time}
P(a) = \frac{\eta \bullet J_a}{\sum_{a^\prime} \eta \bullet J_{a^\prime}}.
\end{equation}
The statistics are thus fully captured by the density vector of the two-time ensemble ($\eta$), and the density vectors corresponding to each outcome of the measurement ($J_a$). Note that the denominator, which ensures that the probabilities sum to one, generally depends upon the choice of measurement. In such cases it is impossible to remove it by merely normalising $\eta$.\footnote{This also means that two density vectors which are identical up to normalization represent the same physical state.} Note that because of this, the probabilities are in general non-linear with respect to the state $\eta$, as well as the measurement $J_a$. 

Finally, we note that we defined density vectors $\eta$ as being mixtures of pure density vectors. We can easily show that this condition is equivalent to asking that the state $\eta$ is ``positive'', in the sense that for any vector $v\otimes v^{\dagger}$ we have $\eta \bullet (v \otimes v^\dagger) \geq 0$ (which is the two-time version of the usual definition of a positive operator: $A$ is positive if, for any $\ket{\psi}$, $\langle \psi |A|\psi\rangle \geq 0$). Hence every density vector $\eta$ is positive and every positive $\eta$ is a density vector. Furthermore, note that the $J$, as defined in \eqref{Krausdensityvector} are positive in the same sense. In particular, this ensures that $\eta$ produces positive probabilities via \eqref{Born2time}.

\begin{figure}[t!]
\begin{center}
\includegraphics[width=\columnwidth]{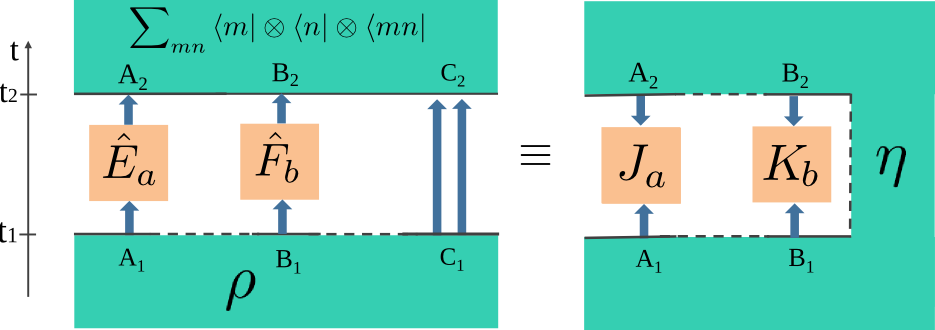}
\caption{Experimental preparation of a mixed bipartite two-time state $\eta$. At time $t_1$, the experimenter prepares the mixed state $\rho = \sum_{ijkl, pqrs} \eta_{ijkl, pqrs}\ket{i}\ket{j}\ket{kl}\bra{p}\bra{q}\bra{rs}\;\in \mathcal{H}^{\vA_1}\otimes \mathcal{H}^{\vB_1}\otimes \mathcal{H}^{\vC_1}$ between Alice, Bob and an ancilla ($C_1$) of dimension $d_{A_2}d_{B_2}$. After Alice and Bob have performed their operations, at  a later time $t_2$,  the experimenter post-selects on the pure state $\sum_{mn}\bra{m}\otimes\bra{n}\otimes\bra{mn}\;\in \mathcal{H}_{\vA_2}\otimes \mathcal{H}_{\vB_2}\otimes \mathcal{H}_{\mathcal{C}_2}$. The effect of this is to have created the bipartite two-time state $\eta$ between times $t_1$ and $t_2$ with the equivalent statistics given by the operators $J_a = E_a \otimes E_a^\dagger$ and $K_b = F_b \otimes F_b^\dagger$. The double arrow on the ancilla indicates that the dimension is greater than that of Alice or Bob's systems.\label{mixedschematic}}
\end{center}
\end{figure}

We will be particularly interested in this paper in bipartite two-time states (shared between Alice and Bob), which can be characterised by $\eta \in \mathcal{H}^{\oA_1}\otimes \mathcal{H}_{\oA_2} \otimes \mathcal{H}^{\oB_1}\otimes \mathcal{H}_{\oB_2}$. (See Fig. \ref{mixedschematic} for the preparation of such a state via post-selection). Given that Alice measures $J_a$ and Bob measures $K_b$, the joint probability to obtain outcomes $a$ and $b$ is given by
\begin{equation}
P(a,b) = \frac{\eta \bullet (J_a\otimes K_b)}{\sum_{a^\prime,b'} \eta \bullet (J_{a^\prime}\otimes K_{b'})}.
\end{equation}
	
\section{The process matrix formalism}

One of the key motivating factors behind the introduction of process matrices in \cite{Oreshkov2012}  was the ability of the formalism to capture ``indefinite causal orders''. We will describe the formalism for two parties; the extension to more parties is straightforward and can be found in \cite{Abbott16}.

The set-up is summarised as follows: Alice and Bob reside in spatially separated laboratories, which are sealed off from the outside world. The doors of Alice's (Bob's) lab may open once to let a system in, and once to let a system out. Within their respective labs Alice and Bob may perform local, quantum mechanical measurements (with the aid of local ancillas, should they require them). Finally, Alice and Bob's labs may be connected in some way. For example it could be the case that on receiving a system, processing it, and then releasing it, Alice passes a quantum system to Bob, who opens his lab doors to let it in (or vice versa). However, they may be connected in completely different ways, corresponding to indefinite causal order, which may not be realisable in quantum mechanics. 

The novel element in this formalism is the so-called ``process matrix" $W$ which specifies how the two labs are connected, and loosely speaking, plays the role of a joint state for Alice and Bob. More precisely, we associate two Hilbert spaces to Alice, one to the system that enters her laboratory, and one to the system that leaves it, and similarly for Bob. The process matrix is a mathematical object playing the role of a generalised ``density operator'' acting on these four Hilbert spaces.

On the other hand, measurements in this formalism are usual quantum measurements. Suppose Alice performs a measurement where  $\hat{E}_{a}^\mu$ denotes the (possibly non-detailed) Kraus operators corresponding to the outcome $a$; the index $\mu$ runs over the various operators corresponding to this individual outcome. In the standard quantum formalism, if the state of Alice's system is $\rho$, the unnormalised state after the measurement, given that the outcome $a$ was obtained, is $\rho_{a}=\mathcal{M}_{a}(\rho) =\sum_\mu \hat{E}_{a}^\mu \rho \, \hat{E}_{a}^{\mu\,\dagger}$, with $\mathcal{M}_{a} : \mathcal{L}(\mathcal{H}^{\vA_1}) \rightarrow \mathcal{L}(\mathcal{H}^{\vA_2} ) $ denoting the linear map \textit{from}  linear operators on the input Hilbert space \textit{to} linear operators on the output space, thus describing the measurement. We similarly denote Bob's Kraus operators by $\hat{F}_{b}^\nu$ and the action of this map as $\mathcal{N}_{b}(\rho) = \sum_\nu \hat{F}_{b}^\nu\rho \,\hat{F}_{b}^{\nu\dagger}$. 

In the process-matrix formalism the measurements of Alice are represented (via a Choi-Jamio\l kowski transformation) by \cite{Oreshkov2012, Araujo1}
\begin{align}\label{e:CJ}
M^{A_1A_2}_{a} &=[\mathcal{I} \otimes \mathcal{M}_{a}(\ket{\Phi^+}\bra{\Phi^+})]^{\mathrm{T}} \nonumber \\
&=\sum_\mu[(\mathbb{I}\otimes \hat{E}_a^\mu)\ket{\Phi^+}\bra{\Phi^+}(\mathbb{I}\otimes \hat{E}_a^{\mu\dagger})]^{\mathrm{T}}
\end{align} 
where $\ket{\Phi^+} = \sum_i \ket{i}\ket{i} \in \mathcal{H}^{\vA_1} \otimes \mathcal{H}^{\vA^\prime_1}$ is the unnormalized maximally entangled state on two copies of Alice's input Hilbert space, and $\mathcal{M}_a: \mathcal{L}(\mathcal{H}^{\vA^\prime_1}) \rightarrow \mathcal{L}(\mathcal{H}^{\vA_2} )$. This state is written in an arbitrary but fixed orthonormal basis $\{\ket{i}\}$ and ${\,\mathrm{T}}$ is the (full) \textit{transpose} operation taken with respect to this basis.  Finally,  
$\mathcal{I} : \mathcal{L}(\mathcal{H}^{\vA_1}) \rightarrow \mathcal{L}(\mathcal{H}^{\vA_1} )$, is the identity map. The measurements of Bob are represented analogously, and denoted $N^{\oB_1\oB_2}_b$.  

Having defined the measurements in this framework the next step is to define the object on which they act. This object is the process matrix $W^{A_1 A_2 B_1 B_2} \in \mathcal{L}(\mathcal{H}^{\vA_1}\otimes \mathcal{H}^{\vA_2} \otimes \mathcal{H}^{\vB_1}\otimes \mathcal{H}^{\vB_2})$, and is an operator acting on the input and output Hilbert spaces of Alice and Bob. The probability rule for the measurement outputs is then given in analogy to the Born rule of quantum mechanics,  
\begin{align} \label{eq:Wprob}
P_W(a,b) &= \tr \left[
W^{A_1 A_2 B_1 B_2} (M^{A_1A_2}_{a} \otimes N^{B_1B_2}_{b})  \right]
.
\end{align}
For $W$ to be a valid process matrix, the probabilities $P_W(a,b)$ must be positive and normalised, $\sum_{ab}P_W(a,b) = 1$, in addition to $W$ itself being a positive operator. These are the only requirements to be a valid process matrix, and the set of valid process matrices is any operator satisfying these constraints.

\section{Connecting process matrices with multi-time states}\label{sec:connect}
We are now in position to present our main result, namely to connect process matrices and multi-time states. In particular, we will show that to every process matrix we can associate a two-time state that will produce the same probabilities for all measurements. This shows that every process matrix can be realised within quantum theory, if both pre- and post-selection are allowed. 

We will present the mapping in two ways; first in terms of matrix elements and second via an isomorphism between Hilbert spaces. Starting with the former, let us denote an arbitrary process matrix as
\begin{align} 
	W^{A_1 A_2 B_1 B_2} &= \sum_{\substack{ijkl\\pqrs}} w_{ijkl, pqrs} \ket{ijkl}\bra{pqrs}, \label{eq:wdec}
\end{align}
\emph{where this decomposition is in the same basis as the $\ket{\Phi^+}$ used to define $M_{a}^{A_1A_2}$ and $M_{b}^{B_1B_2}$}. Then, to this process matrix we associate the bipartite two-time state
\begin{multline}\label{e:etaW}
	\ew = \sum_{\substack{ijkl\\pqrs}} w_{ijkl,pqrs} \; {}_{\vA_2}\!\!\bra{j} \otimes \ket{i}^{\vA_1} \otimes {}_{\vA_1^\dagger}\!\!\bra{p} \otimes \ket{q}^{\vA_2^\dagger} \\\otimes {}_{\vB_2}\!\!\bra{l} \otimes \ket{k}^{\vB_1} \otimes {}_{\vB_1^\dagger}\!\!\bra{r} \otimes \ket{s}^{\vB_2^\dagger}.
\end{multline}
Henceforth, we denote a two-time state which has been mapped from a process matrix $W$ as $\ew$. One can explicitly check that $W$ and $\ew$ lead to the same probabilities for all measurements. We show in appendix \ref{a:connection} that 
\begin{align} 
P_W(a,b) &= \tr \left[W^{A_1 A_2 B_1 B_2} 
			(M^{A_1\,A_2}_{a}
				 \otimes N^{B_1\,B_2}_{b})  
					 \right] \nonumber \\
&= \ew \bullet (J_{a} \otimes K_{b}),
\label{eq:w}
\end{align} 
where $J_{a} = \sum_\mu E_{a}^\mu \otimes E_{a}^{\mu\dagger}$ are the Kraus density vectors of Alice's measurement, and $K_{b} = \sum_\nu F_{b}^\nu \otimes F_{b}^{\nu\dagger}$ of Bob's. 
Since all process matrices satisfy $\sum_{ab}P_W(a,b) = 1$, the above shows that every $\ew$ satisfies 
\begin{equation}
\sum_{ab} \ew \bullet (J_{a}\otimes K_{b}) = 1,
\end{equation}
and hence
\begin{align}\label{e:prob w eta}
P_W(a,b) &= \frac{\ew \bullet (J_{a} \otimes K_{b})}{\sum_{a'b'} \ew \bullet (J_{a'} \otimes K_{b'})} \nonumber \\
&= P_{\ew}(a,b),
\end{align}
which completes the claim, and shows that $W$ and $\ew$ lead to the same probabilities. \\
\indent From this reasoning we also deduce that every $\ew$ has the special property that \emph{the probability rule becomes linear}. 

\begin{figure}[t!]
\begin{center}
\includegraphics[scale=0.45]{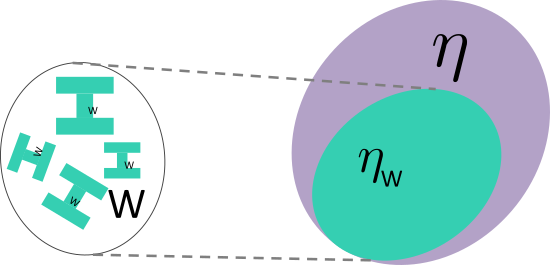}
\caption{Schematic representation of the set of process matrices $\{ W\}$ and their mapping onto a strict subset $\{\ew\}$ of the set of all two-time states $\{\eta \}$. The set $\{\ew\}$ is the set of linear two-time states. 
}
\label{fig:etaSet}
\end{center}
\end{figure}

In  particular, whereas the standard probability rule for two-time states in \eqref{Born2time} is a non-linear function with respect to the density vector $\eta$ and Kraus density vectors $J_{a}$ and $K_{b}$, for those $\ew$ which arise through \eqref{e:etaW} we have $P_{\ew}(a,b) = \ew \bullet (J_{a} \otimes K_{b})$, which is now a linear function of the two-time state, and measurements.
 We will refer to any state that satisfies the property 
\begin{equation}\label{e:linear definition}
P_\eta(a,b) = \eta \bullet (J_{a} \otimes K_{b})
\end{equation}
 as a \emph{linear two-time state}. Note that this is equivalent to $\sum_{ab} \eta \bullet (J_a\otimes K_b) = \eta \bullet (J\otimes K) = 1$. This also highlights the fact that process matrices only map into a subset of two-time states, see Fig. \ref{fig:etaSet}.

We have shown above that any process matrix maps into a linear two-time state. Conversely, every linear two-time state corresponds to some process matrix $W$. In particular, if one writes $\eta$ as in \eqref{e:etaW} and then considers the corresponding matrix $W$ given by \eqref{eq:wdec} then the positivity of $W$ follows from the positivity of $\eta$, and the normalisation of probabilities for any measurement follows from inserting the linearity condition \eqref{e:linear definition} in \eqref{eq:w}, and noting that $P_\eta(a,b)$ is normalised. Thus the matrix $W$ is a process matrix.

We end by giving a second way in which one can express the connection between process matrices and two-time states via a mapping between operators that act on the Hilbert spaces in the process matrix formalism, and vectors in the Hilbert spaces of the two-time state formalism. Recall that $W^{A_1 A_2 B_1 B_2} \in \mathcal{L}(\mathcal{H}^{\vA_1}\otimes \mathcal{H}^{\vA_2} \otimes \mathcal{H}^{\vB_1}\otimes \mathcal{H}^{\vB_2})$ is an operator acting on a four-party Hilbert space. We introduce the following mappings 
\begin{equation}
\begin{split}
	\ket{i}\!\bra{p} 
		\in \mathcal{L}
			 \left( \mathcal{H}^{\vA_1} \right)
			 	 &\rightarrow {}_{\vA_1^\dagger}\!\bra{p} \!\otimes\!  \ket{i}^{\vA_1}
			 	 	\in \mathcal{H}^{\oA_1}, \\ 
	\ket{j}\!\bra{q}
		 \in \mathcal{L} 
		\left( \mathcal{H}^{\vA_2} \right) 
		 	&\rightarrow {}_{\vA_2}\!\bra{j}
		 		 \otimes \ket{q}^{\vA_2^\dagger}
		 		 	 \in \mathcal{H}_{\oA_2}, \\
	\ket{k}\!\bra{r}
		 \in \mathcal{L}
		 	 \left( \mathcal{H}^{\vB_1} \right) 
		 	 	&\rightarrow  {}_{\vB_1^\dagger}\!\bra{r}
		 	 		\otimes   \ket{k}^{\vB_1}
		 	 			\in \mathcal{H}^{\oB_1}, \\
	\ket{l}\!\bra{s}
		\in \mathcal{L} \left( \mathcal{H}^{\vB_2} \right)
			&\rightarrow {}_{\vB_2}\!\bra{l} 
				\otimes \ket{s}^{\vB_2^\dagger}
						\in \mathcal{H}_{B_2}.
\end{split}
\end{equation}
Applying this set of transformations to any process matrix $W$ as in \eqref{eq:wdec} leads to the two-time state $\ew$ as in \eqref{e:etaW}. Note that the second and fourth relations, which map the spaces $\mathcal{L}(\mathcal{H}^{\vA_2})$ and $\mathcal{L}(\mathcal{H}^{\vB_2})$ from the process matrix formalism to the bra-vector spaces and their conjugates, $\mathcal{H}_{\oA_2} \equiv \mathcal{H}_{\vA_2} \otimes \mathcal{H}^{\vA_2^\dagger}$ and $\mathcal{H}_{\oB_2} \equiv \mathcal{H}_{\vB_2} \otimes \mathcal{H}^{\vB_2^\dagger}$ respectively, are basis-dependent mappings, and must be done in the same basis in which the process matrix is represented (i.e.~in this case in the same basis as used in \eqref{eq:wdec}).

\section{The set of two-time states corresponding to process matrices}

In \cite{Oreshkov2012} it is shown that the condition that process matrices give normalised probabilities (that are also positive) for all possible measurements is in fact equivalent to a finite set of conditions, which provide an explicit and compact characterisation of the set of valid process matrices. Here we give a translation of these conditions in the two-time state formalism.

In the bipartite case the set of process matrices  is specified by the following five necessary and sufficient conditions:
\begin{equation} \label{e:W conditions}
	\begin{split}
	W &\ge 0 \\
	\tr W &= d_{A_2}d_{B_2} \\
	{}_{B_1 B_2}W &= {}_{A_2B_1B_2}W, \\
{}_{A_1 A_2}W &= {}_{A_1A_2B_2}W, \\
W &= {}_{A_2}W + {}_{B_2}W - {}_{A_2B_2}W, \\
	\end{split}
\end{equation}
where $d_{A_2}$ ($d_{B_2}$) is the dimension of Alice's (Bob's) output Hilbert space; the notation in the last three equations is defined using the `trace-and-replace' operation
\begin{equation}\label{e:W trace and replace}
{}_X W := \frac{
\mathbb{I}^X}
{d_X 
}\otimes \tr_X W
\,.
\end{equation}

Returning to two-time states, we first need to introduce two operations
\begin{align}\label{e:throw and replace}
\chan{T}{\oA}{\oA'} &:= \frac{1}{\chan{d}{\oA'}{}}
	\mathlarger{\mathbb{I}}^{\mathsmaller{\oA'}}\otimes \mathlarger{\mathbb{I}}_{\mathsmaller{\oA}}, \\
\chan{I}{\oA}{\oA'} &:= \chan{\mathbb{I}}{\vA}{\vA'}
	\otimes
		\chan{\mathbb{I}}{\vA'^\dagger}{\vA^\dagger},
\end{align}
defined in terms of the identity vector \eqref{e:identity op}. The first corresponds to a `throw-away-and-replace' operation, replacing a pre-selected system $\oA$ by the maximally mixed (pre-selected) state on $\oA'$, and is analogous to the `trace-and-replace' operation from \eqref{e:W trace and replace} for process matrices. The second corresponds to the `do-nothing' operation, taking any state of $\oA$ to the same state of $\oA'$. 

Note that in both of these operations we introduce new primed spaces. Depending on the context, these primed spaces should either be thought of as corresponding to a time $t_1'$ just \textit{after} time $t_1$, or to a time $t_2'$ just \emph{before} time $t_2$. For example, $\chan{T}{\oA_1}{\oA_1'}\bullet\et{\oA_1}{}{\oA_2}{} \equiv \ett{\oA_1'}{}{\oA_2}{}{(\eta')} \in \mathcal{H}^{\oA_1'} \otimes \mathcal{H}_{\oA_2}$ is a two-time state between the times $t_1'$ and $t_2$. 

With these operations, a translation of the conditions in \eqref{e:W conditions} is\footnote{Strictly speaking, the third condition from \eqref{e:W conditions} translated into two-time notation reads
\begin{equation}
	(\chan{I}{\oA_2'}{\oA_2} \otimes \chan{T}{\oB_1}{\oB_1^\prime} \otimes \chan{T}{\oB_2^\prime}{\oB_2}) \bullet \ew = 
	(\chan{T}{\oA_2'}{\oA_2} \otimes \chan{T}{\oB_1}{\oB_1^\prime} \otimes \chan{T}{\oB_2^\prime}{\oB_2} ) \bullet \ew,
\end{equation}
which is the same as the third condition in \eqref{e:ew conditions}, with maximally mixed states on Bob's input and output spaces tensored in on both sides of the equation, which does not change the condition. A similar argument holds for the fourth conditions from \eqref{e:W conditions} and \eqref{e:ew conditions}.}
\begin{equation}\label{e:ew conditions}
\begin{split}
	\ew &\geq 0, \\
	\left( \mathlarger{\mathbb{I}}_{\mathsmaller{\oA_1}}\!
	\otimes
		\mathlarger{\mathbb{I}}^{\mathsmaller{\oA_2}}\!
			\otimes
				\mathlarger{\mathbb{I}}_{\mathsmaller{\oB_1}}\!
					\otimes
						\mathlarger{\mathbb{I}}^{\mathsmaller{\oB_2}}\! \right)
							\bullet
								\ew 
									&= d_{\oA_2}d_{\oB_2}, \\
	(\chan{I}{\oA_2'}{\oA_2} 
		\otimes \chan{T}{\oB_1}{\oB_2})
			 \bullet 
			 	\ew &= 
					(\chan{T}{\oA_2'}{\oA_2} 
						\otimes
					 \chan{T}{\oB_1}{\oB_2} )
					 	\bullet 
					 		\ew, \\
	(\chan{T}{\oA_1}{\oA_2} 
		\otimes
			\chan{I}{\oB'_2}{\oB_2} )
			\bullet 
				\ew 
					&=( \chan{T}{\oA_1}{\oA_2} 
						\otimes
						\chan{T}{\oB'_2}{\oB_2} )
							\bullet
								 \ew, \\
	(\chan{I}{\oA_2'}{\oA_2} 
		\otimes
			\chan{I}{\oB_2'}{\oB_2})
				 \bullet 
				 	\ew &= 
				 		(\chan{I}{\oA_2'}{\oA_2}
				 			 \otimes
				 			 	\chan{T}{\oB_2'}{\oB_2})\bullet 
				 			 		\ew \\
									& \quad + ( \chan{T}{\oA_2'}{\oA_2} 
				 			 			\otimes
				 			 			\chan{I}{\oB'_2}{\oB_2})\bullet \ew \\
	&\quad
		- (\chan{T}{\oA_2'}{\oA_2} 
			\otimes
				\chan{T}{\oB'_2}{\oB_2})
					\bullet 	
					\ew.
\end{split}
\end{equation}

\section{Generalised properties of process matrices}\label{sec:newChar}
In this section we show that the properties \eqref{e:ew conditions} of process matrices, that are expressed in terms of the `throw-away-and-replace' and the `do-nothing' channels ($\chan{T}{\oA}{\oA'}$ and $\chan{I}{\oA}{\oA'}$), can be generalized to arbitrary channels. This demonstrates the  usefulness of connecting the process matrix formalism to the formalism of two-time states, and might be useful in future research on process matrices, as it provides further structural information about them. 

In  Appendix \ref{a:condtions} we prove the following result concerning \emph{linear} bipartite two-time states (i.e. about $\ew$ states):\\

\textbf{Theorem} \emph{Given any $\ew$, i.e. any bipartite two time state that satisfies\footnote{Since the overall normalisation of a two-time state is unphysical,  given a two-time state $\eta$ that satisfies $(\chan{J}{\oA_1}{\oA_2} \otimes \chan{K}{\oB_1}{\oB_2}) \bullet \eta= \lambda$ for all $\chan{J}{\oA_1}{\oA_2}$ and $\chan{K}{\oB_1}{\oB_2}$, we can always re-scale $\eta \to \frac{1}{\lambda}\eta$ such that this condition holds.}
\begin{equation}\label{e:linear def}
(\chan{J}{\oA_1}{\oA_2}
	\otimes 
		\chan{K}{\oB_1}{\oB_2})
		\bullet 
			\ew
				= 1,
\end{equation}
for all $\chan{J}{\oA_1}{\oA_2}$ and $\chan{K}{\oB_1}{\oB_2}$ corresponding to completely positive trace preserving maps (i.e. of the form \eqref{Krausdensityvector} and such that $\chan{\mathbb{I}}{\oA_2}{}\bullet \chan{J}{\oA_1}{\oA_2} = \chan{\mathbb{I}}{\oA_1}{}$ and $\chan{\mathbb{I}}{\oB_2}{}\bullet \chan{K}{\oB_1}{\oB_2} = \chan{\mathbb{I}}{\oB_1}{}$) then the following properties hold:
\begin{align}\label{e:linear def consequence}
(\chan{C}{\oA_2'}{\oA_2} \otimes \chan{K}{\oB_1}{\oB_2}) \bullet \ew
			&=(\chan{\tilde{C}}{\oA_2'}{\oA_2} \otimes \chan{K}{\oB_1}{\oB_2}) \bullet \ew, \nonumber\\
(\chan{J}{\oA_1}{\oA_2} \otimes \chan{D}{\oB_2'}{\oB_2}) \bullet \ew
	&= (\chan{J}{\oA_1}{\oA_2} \otimes \chan{\tilde{D}}{\oB_2'}{\oB_2}) \bullet \ew,\nonumber\\
(\chan{C}{\oA_2'}{\oA_2} \otimes \chan{D}{\oB_2'}{\oB_2} ) \bullet \ew &= (\chan{C}{\oA_2'}{\oA_2} \otimes \chan{\tilde{D}}{\oB_2'}{\oB_2} ) \bullet \ew \nonumber\\ 
& \quad + (\chan{\tilde{C}}{\oA_2'}{\oA_2} \otimes \chan{D}{\oB_2'}{\oB_2} ) \bullet \ew \nonumber\\
&\quad -  (\chan{\tilde{C}}{\oA_2'}{\oA_2} \otimes \chan{\tilde{D}}{\oB_2'}{\oB_2} ) \bullet \ew
\end{align}
for all
$\chan{C}{\oA_2'}{\oA_2}$, 
$\chan{\tilde{C}}{\oA_2'}{\oA_2}$, $\chan{D}{\oB_2'}{\oB_2}$ and
$\chan{\tilde{D}}{\oB_2'}{\oB_2}$ corresponding to completely positive trace preserving maps.\footnote{Note that the final condition of \eqref{e:linear def consequence} can be expressed more symmetrically as $[(\chan{C}{\oA_2'}{\oA_2}  -\chan{\tilde{C}}{\oA_2'}{\oA_2})\otimes (\chan{D}{\oB_2'}{\oB_2} - \chan{\tilde{D}}{\oA_2'}{\oA_2})]  \bullet \ew =0$, but we adopt the above form to emphasise its similarity with \eqref{e:W conditions}.}\\
}

We can translate the above conditions into the process matrix formalism, and present a new set of conditions that are satisfied by all process matrices. To do so, let us introduce arbitrary Choi-Jamio\l kowski operators corresponding to completely positive and trace-preserving (CPTP) maps $R^{\oA_2'\oA_2}$, $\tilde{R}^{\oA_2'\oA_2}$ and $M^{\oA_1\oA_2}$ for Alice, and similarly $S^{\oB_2'\oB_2}$, $\tilde{S}^{\oB_2'\oB_2}$ and $N^{\oB_1\oB_2}$ for Bob. Then, the conditions \eqref{e:linear def consequence} can be re-expressed as
\begin{equation}
\begin{split}
\tr&_{\oA_2\oB_1\oB_2}\left[ \left(\mathbb{I}^{\oA_1} \otimes R^{\oA_2'\oA_2}\otimes N^{\oB_1\oB_2} \right) (W\otimes \mathbb{I}^{\oA_2'})\right] \\
&= \tr_{\oA_2\oB_1\oB_2}\left[ \left(\mathbb{I}^{\oA_1}\otimes \tilde{R}^{\oA_2'\oA_2}\otimes N^{\oB_1\oB_2} \right) (W\otimes \mathbb{I}^{\oA_2'})\right], \\
\tr&_{\oA_1\oA_2\oB_2}\left[ \left( M^{\oA_1\oA_2} \otimes \mathbb{I}^{\oB_1} \otimes S^{\oB_2'\oB_2} \right) (W\otimes \mathbb{I}^{\oB_2'})\right] \\
&= \tr_{\oA_1\oA_2\oB_2}\left[ \left( M^{\oA_1\oA_2} \otimes \mathbb{I}^{\oB_1} \otimes \tilde{S}^{\oB_2'\oB_2} \right) (W\otimes \mathbb{I}^{\oB_2'})\right], \\
\tr&_{\oA_2\oB_2}\left[ \left(\mathbb{I}^{\oA_1\oB_1} \otimes R^{\oA_2'\oA_2} \otimes S^{\oB_2'\oB_2} \right) (W\otimes \mathbb{I}^{\oA_2'\oB_2'})\right] \\
&=\tr_{\oA_2\oB_2}\left[ \left(\mathbb{I}^{\oA_1\oB_1} \otimes R^{\oA_2'\oA_2} \otimes \tilde{S}^{\oB_2'\oB_2} \right) (W\otimes \mathbb{I}^{\oA_2'\oB_2'})\right] \\
&\quad+\tr_{\oA_2\oB_2}\left[ \left(\mathbb{I}^{\oA_1\oB_1} \otimes \tilde{R}^{\oA_2'\oA_2} \otimes S^{\oB_2'\oB_2} \right) (W\otimes \mathbb{I}^{\oA_2'\oB_2'})\right] \\
&\quad-\tr_{\oA_2\oB_2}\left[ \left(\mathbb{I}^{\oA_1\oB_1} \otimes \tilde{R}^{\oA_2'\oA_2} \otimes \tilde{S}^{\oB_2'\oB_2} \right) (W\otimes \mathbb{I}^{\oA_2'\oB_2'})\right].
\end{split}
\end{equation}
These conditions are not directly implied by the conditions \eqref{e:W conditions}, and may prove of independent interest for proving new results regarding process matrices. 

\begin{figure}[t!]
\begin{center}
\includegraphics[scale=0.38]{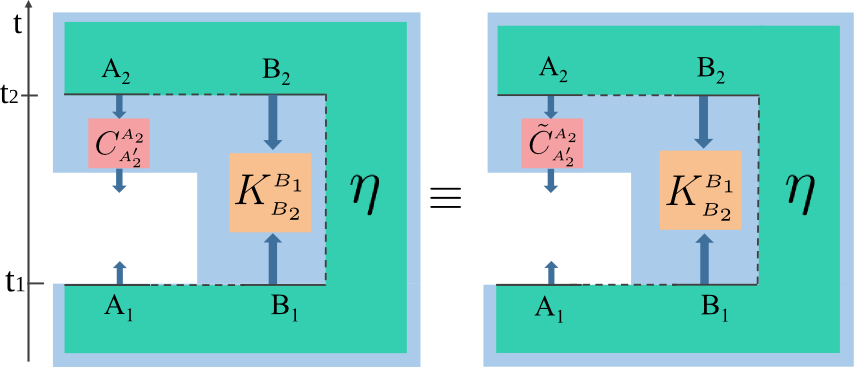}
\caption{Graphical representation of the first condition in \eqref{e:linear def consequence}.}
\end{center}
\end{figure}

\section {Indefinite causal structure versus indefinite temporal structure.}
In the usual causal scenario an event A can be a cause of an event B only if A takes place before B, and B can be a cause of A only if B takes place first. The most general causal structure is a simple mixture of these cases, by probabilistically sometimes letting A be before B and sometimes B before A.  In quantum mechanics one can envisage a more interesting situation in which one can arrange for a {\it superposition} of these two time orderings, resulting in an ``indefinite'' causal order.  The process matrix formalism is inspired by this idea.

Note that in the above the causal structure is constrained by the temporal structure, and indefinite causal order is induced by indefinite temporal order. On the other hand, given the correspondence between two-time states and process matrices presented here, we have implicitly shown that in quantum mechanics with postselection (and presumably in every post-selected probabilistic theory) temporal order and causal order are independent notions. In particular one can have indefinite causal order even with definite temporal order. 

In the process matrix formalism the actual time at which Alice's and Bob's actions take place is left undefined. The only ordering we have is a local time order inside Alice and Bob’s labs, and nothing is said about the relative timing between Alice and Bob. The possibility is left open for Alice to pass her output state to Bob as his input state, and for Bob to pass his output state to Alice as her input state, in some sort of potential superposition, though how exactly this is encoded in the formalism is less clear. 

On the other hand, in the two-state formalism the temporal order is perfectly well-defined: pre-selection takes place at time $t_1$, post-selection at time $t_2$, and the actions of both Alice and Bob take place between these times. Since their initial times are synchronised, in the usual causal scenario, there would be no way for Alice to send her final state to Bob, or vice-versa. However, using post-selection ``effective” communication'' can be realised using the entanglement present in the joint two-time state, which enables, for example, Alice's final state to be teleported backward in time to Bob’s initial time. Thus we can realise a causal order from Alice to Bob, or vice versa, or even an indefinite causal order (when the joint two-time state corresponds to a process matrix with indefinite causal order), despite having definite temporal order.\footnote{Note that in fact any fixed timings for the pre- and post-selections of Alice and Bob, not necessarily the synchronised timings considered above, allows for indefinite causal order with definite temporal order.}

\section{Discussion}
In our paper we addressed the question of the relation between the process matrix formalism and quantum theory: does the process matrix formalism imply new phenomena, not present in quantum theory or not? We showed that correlations predicted by the process matrix formalism are identical to those obtainable in a subset of pre- and post-selected quantum states. In this sense, the process matrix formalism is contained within quantum theory. The subtle question is that of the probability with which we can prepare situations described by process matrices. In ordinary quantum theory pre- and post-selected states cannot be prepared with certainty, unless they are trivial (i.e. no post-selection): as their name suggests, sometimes we don't succeed in obtaining them, so, at the end of the experiment we need to select the cases when we were successful and reject the other cases. Hence, if Nature would turn out to be such that all process matrices can be prepared with certainty, then Nature may not be described by  quantum theory, but something new. (Here we say ``may not" rather than ``would not", since we have not proven that trivial post-selections are not enough, or that quantum mechanics does not allow simulating W matrices by some other means than by pre- and post-selection.) On the other hand, if one enlarges the scope of quantum theory to quantum theory + fundamental post-selection, in which nature provides the ``post-selected" state with certainty  (such as by giving a final state of the universe, in addition to and independent of the initial state), then the situations described by the process matrix formalism are completely contained within it.

As noted above, the correlations described by the process matrix formalism are the same as those arising from a {\it subset} of pre- and post-selected states. This raises a new and important question: Why only this subset? Why not other pre- and post-selected states? As we have seen, the limitation to this set of states stems from the requirement that the probabilities are linear functions of the states and of the measurements. What would be problematic if the probabilities were non-linear functions? After all, in the context of pre- and post selection the probabilities are non-linear in general. We will address this question in forthcoming work. 

We generalised the defining conditions of the process matrices which are given by relations that are obeyed when the systems are subjected to simple ``trace and replace” channels to arbitrary channels. This may be useful in proving further results, or to allow us to gain a deeper understanding of process matrices.

We also discussed the issues of indefinite temporal and causal order and argued that they are actually two independent concepts.

Finally, we note a technical issue which stems from a deeper conceptual issue and shows an advantage of viewing process matrix situations within the two-time state formalism. The process matrix formalism, which is based on the Choi-Jamio\l kowski representation, and in which process-matrices are like density operators on standard ket vector spaces, is basis dependent. In particular, this leads to the rather complicated relation \eqref{e:CJ} between the Kraus operators that describe Alice and Bob’s measurements and the corresponding operators $M_a^{\oA_1\oA_2}$ and $N_b^{\oB_1\oB_2}$ that are used for calculating probabilities. On the other hand, the two-time formalism, in which a ket vector and a bra vector space are used to represent the forward and backward in time propagating states respectively  (and vice versa for the daggered spaces), is basis independent. Indeed, the state $\sum_i |i\rangle^{\mathcal{A}_1} |i\rangle^{\mathcal{A}_2}$ is basis dependent, while the two-time equivalent  $\sum_i |i\rangle^{\mathcal{A}_1} \otimes  \,_{\mathcal{A}_2} \langle i|$ is basis independent. Correspondingly, the relation \eqref{Krausdensityvector} between the Kraus operators $E_a$ and the operator $J_a$  used for calculating probabilities is straightforward.

\section*{Acknowledgements}
YG acknowledges funding from the Austrian Science Fund (FWF) through the START project Y879-N27. SP and PS acknowledge support from the ERC through the AdG NLST. PS acknowledges support from the Royal Society. RS and NB acknowledge financial support from the Swiss National Science Foundation (Starting grants DIAQ and QSIT). AJS acknowledges support from the FQXi via the SVCF.

\onecolumngrid
\bibliography{processRefs}

\newpage

\section*{Appendix}
\begin{appendix}

\section{Proof of the connection between process matrices and two-time states} 
Here we give the proof  of \eqref{eq:w}, which connects  $ P_W(a,b)$ and $\ew$.

	\begin{align} \label{a:connection}
	P_W(a,b) &= \tr \left[W^{A_1 A_2 B_1 B_2} (M^{A_1\,A_2}_{a} \otimes N^{B_1\,B_2}_{b}) \right] \nonumber \\
	&= \tr \sum_{\mu\nu}W^{A_1 A_2 B_1 B_2} \left(\left[(\mathbb{I} \otimes \hat{E}^\mu_{a})
\ket{\Phi^+}\bra{\Phi^+}(\mathbb{I} \otimes \hat{E}^{\mu\dagger}_{a})\right]^{\mathrm{T}}\otimes\left[(\mathbb{I} \otimes \hat{F}^\nu_{b})
\ket{\Phi^+}\bra{\Phi^+}(\mathbb{I} \otimes \hat{F}^{\nu\dagger}_{b})\right]^{\mathrm{T}} \right) \nonumber\\
	&= \tr \sum_{\mu\nu}[W^{A_1 A_2 B_1 B_2}]^\mathrm{T} \left((\mathbb{I} \otimes \hat{E}^\mu_{a})\ket{\Phi^+}\bra{\Phi^+}(\mathbb{I} \otimes \hat{E}^{\mu\dagger}_{a})\otimes(\mathbb{I} \otimes \hat{F}^\nu_{b})
\ket{\Phi^+}\bra{\Phi^+}(\mathbb{I} \otimes \hat{F}^{\nu\dagger}_{b}) \right) \nonumber\\
	&=\tr\sum_{\substack{ijkl\\pqrs}} 
		\sum_{\mu\nu}
			 \sum_{tuvw} 
			 	 w_{ijkl, pqrs}
			 	 	 \ket{pqrs}\bra{ijkl} 
			 	 	 	 \left(\ketbra{t}{u} 
			 	 	 	 	\otimes
			 	 	 	 		  \hat{E}_{a}^{\mu} \ketbra{t}{u} \hat{E}_{a}^{\mu\dagger} 
			 	 	 	 		  	\otimes
			 	 	 	 		  		\ketbra{v}{w}
			 	 	 	 		  			 \otimes  \hat{F}_{b}^{\nu} 
			 	 	 	 		  			 	\ketbra{v}{w} \hat{F}_{b}^{\nu\dagger} \right) \nonumber\\
	&= \sum_{\substack{ijkl\\pqrs}} 
		\sum_{\mu\nu} 
			 w_{ijkl, pqrs}  \, \bra{j}
			 	  \hat{E}_{a}^{\mu}
			 	  	 \ketbra{i}{p} 
			 	  	 	\hat{E}_{a}^{\mu\dagger} 
			 	  	 		\ket{q} \bra{l}
			 	  	 			 \hat{F}_{b}^{\nu}
			 	  	 			 	 \ketbra{k}{r} 
			 	  	 			 	 	\hat{F}_{b}^{\nu\dagger} \ket{s} \nonumber \\
	&= \left(\sum_{\substack{ijkl\\pqrs}} 
		w_{ijkl,pqrs} \; {}_{\vA_2}\!\!\bra{j}\! 
			\otimes \!
				\ket{i}^{\vA_1} \!\!
					\otimes\!\!
						{}_{\vA_1^\dagger}\!\!\bra{p}\! 
							\otimes\! \ket{q}^{\vA_2^\dagger}\! 
								\otimes \!{}_{\vB_2}\!\!\bra{l} 
									\!\otimes \!\ket{k}^{\vB_1} 
										\!\!\otimes 
											\!{}_{\vB_1^\dagger}
												\!\!\bra{r}\!
													 \otimes\! \ket{s}^{\vB_2^\dagger}\right)\bullet \left(\sum_\mu 
				 	E_{a}^\mu 
				 		\otimes E_{a}^{\mu\dagger} 
				 			\otimes \sum_\nu F_{b}^\nu
				 				 \otimes F_{b}^{\nu\dagger}\right) \nonumber \\
	&= \ew \bullet (J_{a} \otimes K_{b}).
	\end{align}

\section{Proof of the generalized properties of process matrices} \label{a:condtions} 
In this appendix we give the proof of the generalized properties of process matrices given in the main text \eqref{e:linear def consequence}. We begin by restating the theorem:

\textbf{Theorem} \emph{Given any $\ew$, i.e. any bipartite two time state that satisfies
\begin{equation}\label{ae:linear def}
(\chan{J}{\oA_1}{\oA_2}
	\otimes 
		\chan{K}{\oB_1}{\oB_2})
		\bullet 
			\ew
				= 1,
\end{equation}
for all $\chan{J}{\oA_1}{\oA_2}$ and $\chan{K}{\oB_1}{\oB_2}$ corresponding to completely positive trace preserving maps (i.e. of the form \eqref{Krausdensityvector} and such that $\chan{\mathbb{I}}{\oA_2}{}\bullet \chan{J}{\oA_1}{\oA_2} = \chan{\mathbb{I}}{\oA_1}{}$ and $\chan{\mathbb{I}}{\oB_2}{}\bullet \chan{K}{\oB_1}{\oB_2} = \chan{\mathbb{I}}{\oB_1}{}$) then the following properties hold:
\begin{align}\label{ae:linear def consequence}
(\chan{C}{\oA_2'}{\oA_2} \otimes \chan{K}{\oB_1}{\oB_2}) \bullet \ew
			&=(\chan{\tilde{C}}{\oA_2'}{\oA_2} \otimes \chan{K}{\oB_1}{\oB_2}) \bullet \ew, \nonumber\\
(\chan{J}{\oA_1}{\oA_2} \otimes \chan{D}{\oB_2'}{\oB_2}) \bullet \ew
	&= (\chan{J}{\oA_1}{\oA_2} \otimes \chan{\tilde{D}}{\oB_2'}{\oB_2}) \bullet \ew,\nonumber\\
(\chan{C}{\oA_2'}{\oA_2} \otimes \chan{D}{\oB_2'}{\oB_2} ) \bullet \ew &= (\chan{C}{\oA_2'}{\oA_2} \otimes \chan{\tilde{D}}{\oB_2'}{\oB_2} ) \bullet \ew  + (\chan{\tilde{C}}{\oA_2'}{\oA_2} \otimes \chan{D}{\oB_2'}{\oB_2} ) \bullet \ew -  (\chan{\tilde{C}}{\oA_2'}{\oA_2} \otimes \chan{\tilde{D}}{\oB_2'}{\oB_2} ) \bullet \ew
\end{align}
for all
$\chan{C}{\oA_2'}{\oA_2}$, 
$\chan{\tilde{C}}{\oA_2'}{\oA_2}$, $\chan{D}{\oB_2'}{\oB_2}$ and 
$\chan{\tilde{D}}{\oB_2'}{\oB_2}$ corresponding to completely positive trace preserving maps.
}

\textbf{Proof: } Consider the following CPTP map for Alice
\begin{equation}\label{e:J proof}
\chan{\tilde{J}}{\oA_1}{\oA_2} = \chan{T}{\oA_1}{\oA_2} + \epsilon(\chan{C}{\oA_2'}{\oA_2}
	 - \chan{\tilde{C}}{\oA_2'}{\oA_2})\bullet \chan{X}{\oA_1}{\oA_2'}
\end{equation}
where $\chan{T}{\oA_1}{\oA_2}$ is the throw-away-and-replace operation (c.f. \eqref{e:throw and replace}), $\chan{C}{\oA_2'}{\oA_2}$ and $\chan{\tilde{C}}{\oA_2'}{\oA_2}$ correspond to two arbitrary CPTP maps,  $\chan{X}{\oA_1}{\oA_2'} \in \mathcal{H}_{\oA_1} \otimes \mathcal{H}^{\oA_2'}$ is an arbitrary vector\footnote{The only requirement on $\chan{X}{\oA_1}{\oA_2'}$ is that $\chan{X}{\oA_1}{\oA_2'} \bullet \et{\oA_1}{}{\oA_2'}{} \in \mathbb{R}$ for all $\et{\oA_1}{}{\oA_2'}{} \in \mathcal{H}^{\oA_1}\otimes \mathcal{H}_{\oA_2'}$, which can be seen as a `hermiticity' requirement.}, and $\epsilon > 0$ is a positive constant, taken to be sufficiently small such that $\chan{\tilde{J}}{\oA_1}{\oA_2}$ is positive.  Similarly, for Bob, 
\begin{equation}\label{e:K proof}
\chan{\tilde{K}}{\oB_1}{\oB_2} = \chan{T}{\oB_1}{\oB_2} + \delta(\chan{D}{\oB_2'}{\oB_2}
	 - \chan{\tilde{D}}{\oB_2'}{\oB_2})\bullet \chan{Y}{\oB_1}{\oB_2'}
\end{equation}
where $\chan{D}{\oA_2'}{\oA_2}$ and $\chan{\tilde{D}}{\oA_2'}{\oA_2}$ are two arbitrary CPTP maps,  $\chan{Y}{\oB_1}{\oB_2'} \in \mathcal{H}_{\oB_1} \otimes \mathcal{H}^{\oB_2'}$ is an arbitrary vector, and $\delta > 0$ is a sufficiently small positive constant so that $\chan{\tilde{K}}{\oB_1}{\oB_2}$ is positive.

Now, for linear two-time states it follows that
\begin{equation}
(\chan{\tilde{J}}{\oA_1}{\oA_2}\otimes \chan{K}{\oB_1}{\oB_2}) \bullet \et{\oA_1}{\oB_1}{\oA_2}{\oB_2} = (\chan{T}{\oA_1}{\oA_2}\otimes \chan{K}{\oB_1}{\oB_2}) \bullet \et{\oA_1}{\oB_1}{\oA_2}{\oB_2},
\end{equation}
for all CPTP maps $\chan{K}{\oB_1}{\oB_2}$. Using \eqref{e:J proof} to expand out $\chan{\tilde{J}}{\oA_1}{\oA_2}$, this is seen to be equivalent to
\begin{equation}
\left[\left((\chan{C}{\oA_2'}{\oA_2}-\chan{\tilde{C}}{\oA_2'}{\oA_2})\bullet \chan{X}{\oA_1}{\oA_2'}\right) \otimes \chan{K}{\oB_1}{\oB_2} \right]\bullet \et{\oA_1}{\oB_1}{\oA_2}{\oB_2} = 0,
\end{equation}
where we have used the fact that $\epsilon > 0$ to cancel it. The only way that this can hold for an arbitrary vector $\chan{X}{\oA_1}{\oA_2'}$ is if
\begin{equation}
(\chan{C}{\oA_2'}{\oA_2}
	\otimes 
		\chan{K}{\oB_1}{\oB_2})
			\bullet 
			\et{\oA_1}{\oB_1}{\oA_2}{\oB_2}=(\chan{\tilde{C}}{\oA_2'}{\oA_2}
	\otimes 
		\chan{K}{\oB_1}{\oB_2})
			\bullet 
			\et{\oA_1}{\oB_1}{\oA_2}{\oB_2}
\end{equation}
which is the first condition of the theorem. 

An identical argument can be made, by starting from $(\chan{J}{\oA_1}{\oA_2}\otimes \chan{\tilde{K}}{\oB_1}{\oB_2}) \bullet \et{\oA_1}{\oB_1}{\oA_2}{\oB_2} = (\chan{J}{\oA_1}{\oA_2}\otimes \chan{T}{\oB_1}{\oB_2}) \bullet \et{\oA_1}{\oB_1}{\oA_2}{\oB_2}$, where $\chan{J}{\oA_1}{\oA_2}$ is an arbitrary CPTP map, which leads to the second result of the theorem, 
\begin{equation}
(\chan{J}{\oA_1}{\oA_2}
	\otimes 
	\chan{D}{\oB_2'}{\oB_2}
		) 
	\bullet 
	\et{\oA_1}{\oB_1}{\oA_2}{\oB_2}=
(\chan{J}{\oA_1}{\oA_2}
	\otimes 
	\chan{\tilde{D}}{\oB_2'}{\oB_2}	
		) 
	\bullet 
	\et{\oA_1}{\oB_1}{\oA_2}{\oB_2}
\end{equation}
Finally, by using the relation
\begin{equation}
(\chan{\tilde{J}}{\oA_1}{\oA_2}\otimes \chan{\tilde{K}}{\oB_1}{\oB_2}) \bullet \et{\oA_1}{\oB_1}{\oA_2}{\oB_2} = (\chan{T}{\oA_1}{\oA_2}\otimes \chan{T}{\oB_1}{\oB_2}) \bullet \et{\oA_1}{\oB_1}{\oA_2}{\oB_2},
\end{equation}
it follows that
\begin{equation}
\left((\chan{C}{\oA_2'}{\oA_2}-\chan{\tilde{C}}{\oA_2'}{\oA_2})\bullet \chan{X}{\oA_1}{\oA_2'}\right) \otimes\left( (\chan{D}{\oB_2'}{\oB_2}-\chan{\tilde{D}}{\oB_2'}{\oB_2})\bullet \chan{Y}{\oB_1}{\oB_2'}\right) \bullet \et{\oA_1}{\oB_1}{\oA_2}{\oB_2} = 0,
\end{equation}
having used the two results already proved. Since this must hold for all $\chan{X}{\oA_1}{\oA_2'}$ and $\chan{Y}{\oB_1}{\oB_2'}$ it follows that
\begin{align}
(\chan{C}{\oA_2'}{\oA_2} \otimes \chan{D}{\oB_2'}{\oB_2} ) \bullet \et{\oA_1}{\oB_1}{\oA_2}{\oB_2} &= (\chan{C}{\oA_2'}{\oA_2} \otimes \chan{\tilde{D}}{\oB_2'}{\oB_2} ) \bullet \et{\oA_1}{\oB_1}{\oA_2}{\oB_2} + (\chan{\tilde{C}}{\oA_2'}{\oA_2} \otimes \chan{D}{\oB_2'}{\oB_2} ) \bullet \et{\oA_1}{\oB_1}{\oA_2}{\oB_2} -  (\chan{\tilde{C}}{\oA_2'}{\oA_2} \otimes \chan{\tilde{D}}{\oB_2'}{\oB_2} ) \bullet \et{\oA_1}{\oB_1}{\oA_2}{\oB_2}
\end{align}
which completes the proof. 
\end{appendix}

\end{document}